\documentclass[epj]{svjour}
\usepackage{graphics}
\begin{document}
\title{Electroweak Penguin $B$ Decays at Belle}
\author{Patrick Koppenburg, for the Belle Collaboration
% \thanks{\emph{Present address:} Insert the address here if needed}%
}                     % Do not remove
\institute{KEK --- High Energy Accelerator Research Organization, Tsukuba, Japan}
\date{Received: date / Revised version: date}
\abstract{
We summarise the most recent results of the Belle
experiment about flavour changing neutral current 
(FCNC) radiative and (semi-) leptonic $B$ decays. 
In particular, we report about the first observation of the
decays $B\to K^\ast \ell^+\ell^-$, $B\to \phi K \gamma$,
the inclusive $B\to X_s \ell^+\ell^-$. We also report about
searches for $B\to \ell^+\ell^-$ decay and for 
$CP$ asymmetries in $B\to K^\ast \gamma$.
 % end of PACS codes
} %end of abstract
\maketitle
%
%%%%%%%%%%%%%%%%%%%%%%%%%%%%%%%%%%%%%%%%%%%%%%%%%%%%%%%%%%%%%%%%%%%%%%%%%%%%%%%%
%
\section{Introduction}\label{intro}
%
%%%%%%%%%%%%%%%%%%%%%%%%%%%%%%%%%%%%%%%%%%%%%%%%%%%%%%%%%%%%%%%%%%%%%%%%%%%%%%%%
Since the first observation of a penguin decay ten years ago~\cite{Ref:CLEOkg},
radiative $B$ decays have been a powerful tool to constrain physics beyond the
Standard Model. Today we enter an era of precision measurements as
the error on the $B\to K^\ast\gamma$ branching fraction 
is about to become systematics-dominated and as
we start to observe more rare decays like $b\to s\bar{s}s\gamma$. 
In the future $b\to s\gamma$ transitions may be used to probe the
kinematic properties the $B$ decays, which is useful to understand the $V_{ub}$
extraction from semileptonic decays, and may also provide a handle on $V_{td}$
once the Cabbibo-suppressed $b\to d\gamma$ decays are seen.

At the price of an additional suppression by $\alpha_{\rm e.m.}$, one gets
flavour-changing neutral current (FCNC) semileptonic $b\to s\ell\ell$ decays,
where the lepton pair provides other observables, like 
the forward-backward charge asymmetry, which are much more powerful
to constrain the Standard Model and its extensions.

In this report we summarise the latest results from Belle~\cite{Ref:Belle}
about the above mentioned decays and also about purely leptonic $B\to\ell\ell$
decays.

%%%%%%%%%%%%%%%%%%%%%%%%%%%%%%%%%%%%%%%%%%%%%%%%%%%%%%%%%%%%%%%%%%%%%%%%%%%%%%%%
%
\section{Radiative decays}\label{sec:gamma}
%
%%%%%%%%%%%%%%%%%%%%%%%%%%%%%%%%%%%%%%%%%%%%%%%%%%%%%%%%%%%%%%%%%%%%%%%%%%%%%%%%
While we start to perform precise branching fraction and  
$CP$ asymmetry measurements in the $B\to K^\ast\gamma$ decay,
which cannot be considered as ``rare'' at $B$ factories anymore, 
most of the partial width of $B\to X_s \gamma$ is yet still unknown.
Thus the search for more exclusive final states is needed to 
achieve a better understanding of the hadronic structure of this decay.
%%%%%%%%%%%%%%%%%%%%%%%%%%%%%%%%%%%%%%%%%%%%%%%%%%%%%%%%%%%%%%%%%%%%%%%%%%%%%%%%
\subsection{First observation $B\to K\phi\gamma$}\label{sec:kphig}
Using $90\:\rm fb^{-1}$, we observe the decay $B^-\to\phi K^-\gamma$~\cite{Ref:Kphig}.
This is the first observation of a radiative $b\to s\bar{s}s\gamma $ process.
The decay is reconstructed using a high-energy photon, two
oppositely charged kaons required to form
the $\phi$ mass within $10\:\rm MeV$ ($\sim 3\sigma$),
and one additional $K^-$ or $K_S^0$.
We observe $21.6\pm 5.6$ events in the charged mode, 
(corresponding to a statistical significance of $5.5\sigma$),
and $5.8\pm 3.0$ events in the neutral mode ($3.3\sigma$).
The preliminary measured branching fractions are:
\begin{eqnarray*}
{\mathcal B}\left(B^-\to K^-\phi\gamma\right) 
  & = & \left(3.4\pm0.9\pm0.4\right)\cdot 10^{-6} \\
{\mathcal B}\left(B^0\to K^0\phi\gamma\right) 
  & = & \left(4.6\pm2.4\pm0.6\right)\cdot 10^{-6}.
\end{eqnarray*}
In the latter mode we also give an upper limit for the branching fraction
at $8.3\cdot 10^{-6}$ at $90$\% confidence level. 

%%%%%%%%%%%%%%%%%%%%%%%%%%%%%%%%%%%%%%%%
\begin{figure}
\resizebox{0.48\columnwidth}{!}{\includegraphics{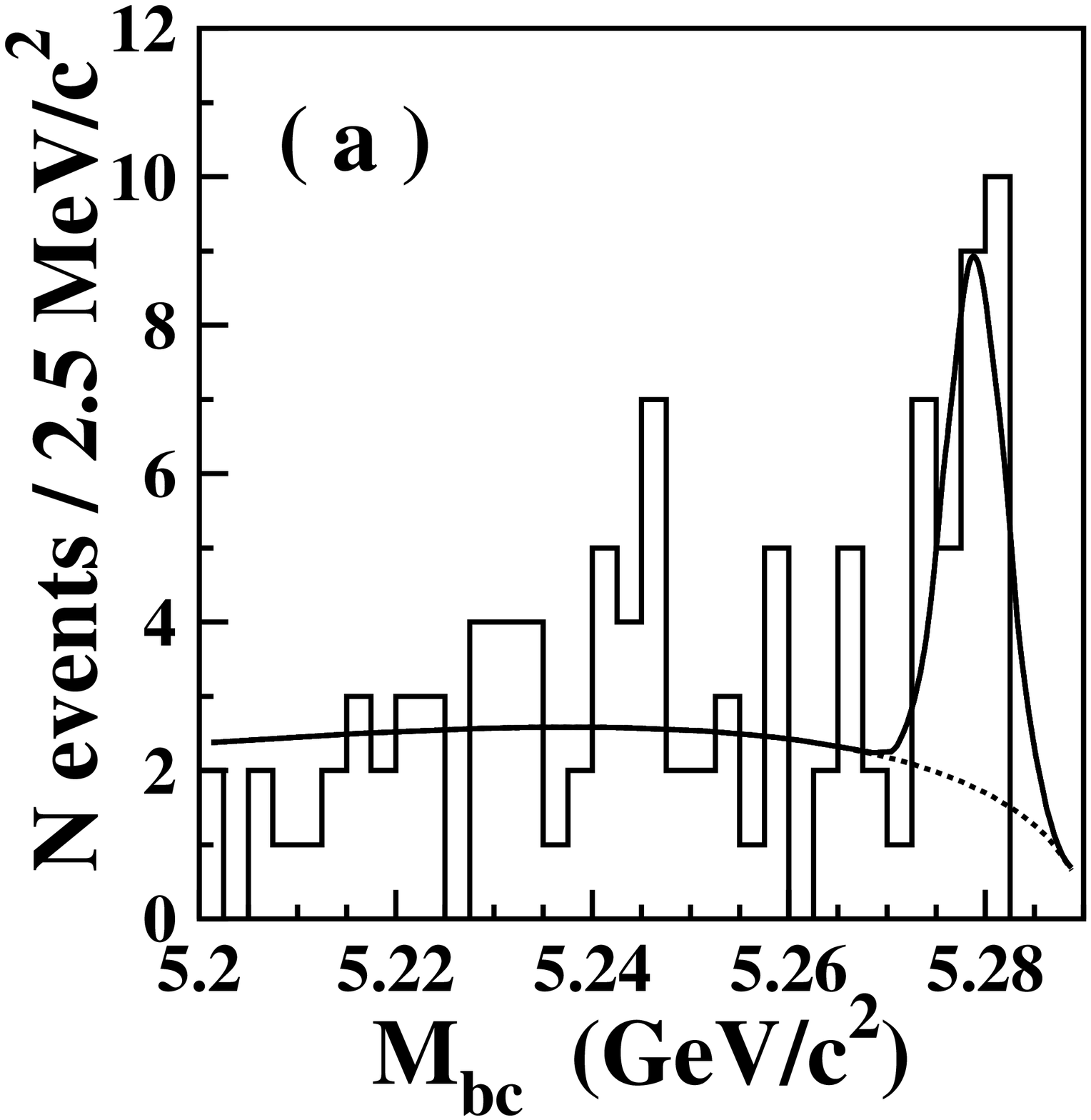}}
\resizebox{0.48\columnwidth}{!}{\includegraphics{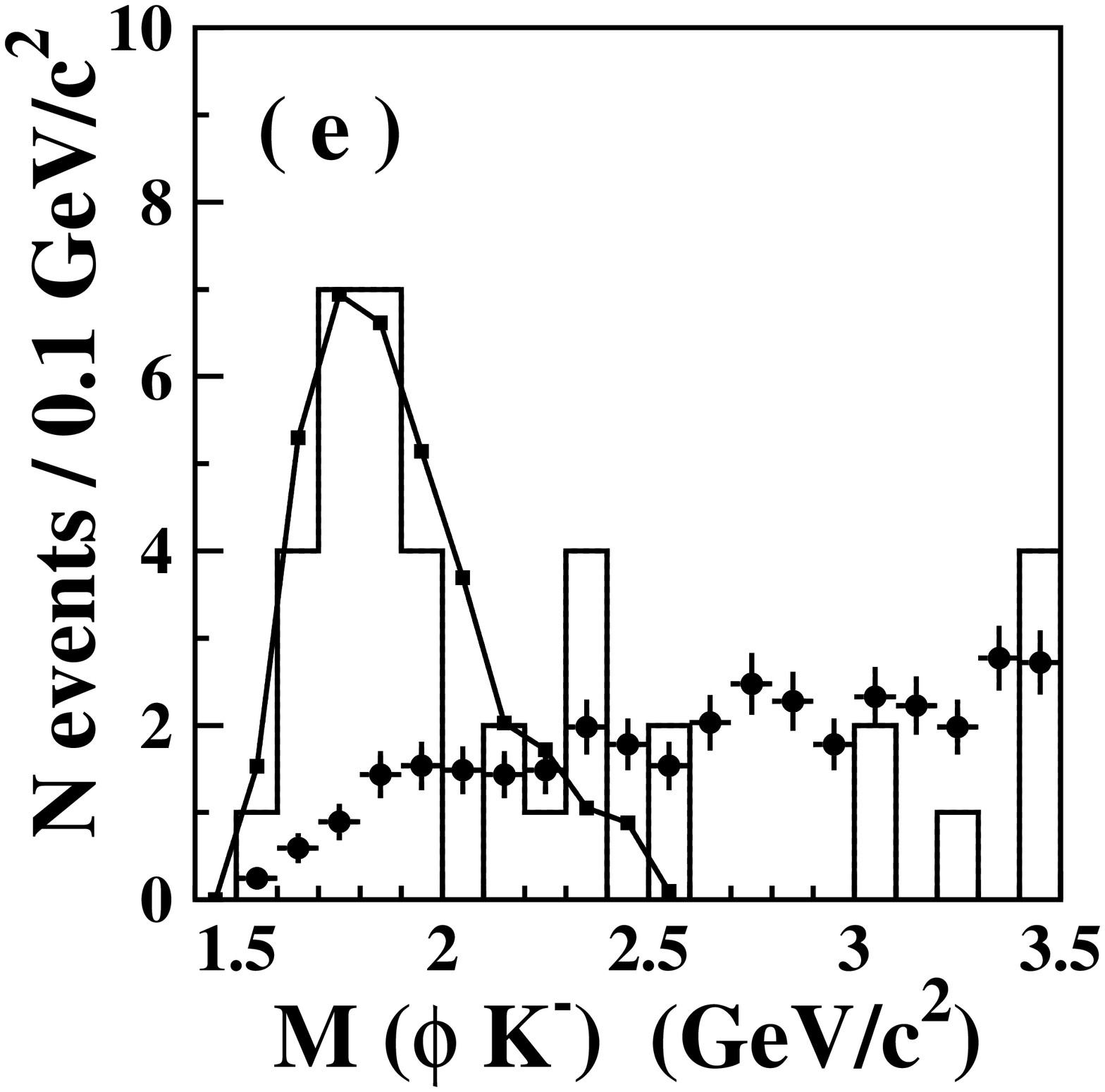}}
\caption{$m_{\rm bc}$ fit (left) 
         and $m_{\phi K}$ (right) 
	 for $K\phi\gamma$ final state. 
	 The measured (solid) $m_{\phi K}$ distribution 
	 is compared to MC simulations basing on a phase-space
	 model (circles) or adjusted to follow the data 
	 (squares connected by a line).} 
\label{fig:Kphig}  
\end{figure}
%%%%%%%%%%%%%%%%%%%%%%%%%%%%%%%%%%%%%%%%
The beam-constrained mass fit for the charged mode is
shown in Fig.~\ref{fig:Kphig} (left). The right hand side figure
shows that the $\phi K^-$ mass distribution differs from a naive
three-body phase-space decay model. Yet the low statistics
do not allow to draw any conclusion about the structure.

%%%%%%%%%%%%%%%%%%%%%%%%%%%%%%%%%%%%%%%%%%%%%%%%%%%%%%%%%%%%%%%%%%%%%%%%%%%%%%%%
\subsection{$CP$ asymmetry in $B\to K^\ast\gamma$}\label{sec:kg}
%%%%%%%%%%%%%%%%%%%%%%%%%%%%%%%%%%%%%%%%%%%%%%%%%%%%%%%%%%%%%%%%%%%%%%%%%%%%%%%%
Among radiative penguin decays, the
$B\to K^\ast\gamma$ decay allows the most precise measurements.
We observe 700 such decays~\cite{Ref:Kgamma}, using a $78\:\rm fb^{-1}$ 
data sample and reconstructing the $K^\ast$ in all visible
final states $K^+\pi^-$, $K_S^0\pi^0$, 
$K^+\pi^0$, $K_S^0\pi^+$
(charge conjugation is implied
throughout this report except where mentioned).
%%%%%%%%%%%%%%%%%%%%%%%%%%%%%%%%%%%%%%%%
\begin{figure}
\resizebox{\columnwidth}{!}{\includegraphics{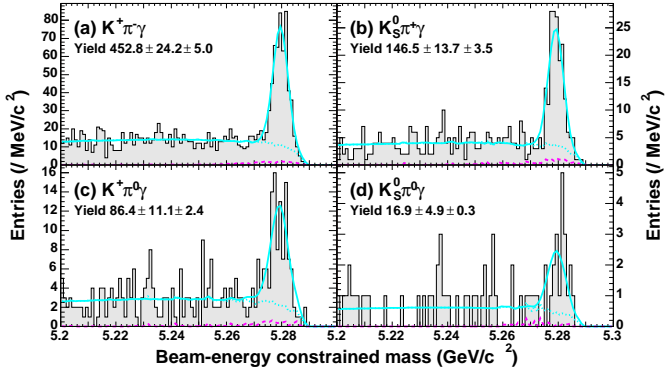}}
\caption{Beam-constrained mass fits for $K^\ast\gamma$ final states.}
\label{fig:Kstg}  
\end{figure}
%%%%%%%%%%%%%%%%%%%%%%%%%%%%%%%%%%%%%%%%
The corresponding beam-constrained mass ($m_{\rm bc}$) distributions
are shown in Fig.~\ref{fig:Kstg}.
The preliminary branching fractions are found to be
\begin{eqnarray*}
{\mathcal B}\left(B^0\to K^{\ast 0}\gamma\right) & = & \left(4.09\pm0.21\pm0.19\right)\cdot 10^{-5} \\
{\mathcal B}\left(B^+\to K^{\ast +}\gamma\right) & = & \left(4.40\pm0.33\pm0.24\right)\cdot 10^{-5},
\end{eqnarray*}
where the first error is statistical and the second systematic.
Fitting the event yields separately for the two flavour eigenstates of the $B$
meson (thus excluding the $K_S^0\pi^0\gamma$ final state) we get a 
measurement of the $CP$ asymmetry:
$$
  A_{CP}\left(B\to K^\ast\gamma\right) = -0.001 \pm 0.044 \pm 0.008.
$$
%%%%%%%%%%%%%%%%%%%%%%%%%%%%%%%%%%%%%%%%%%%%%%%%%%%%%%%%%%%%%%%%%%%%%%%%%%%%%%%%
%
\section{Semileptonic Penguin decays}\label{sec:semil}
%
%%%%%%%%%%%%%%%%%%%%%%%%%%%%%%%%%%%%%%%%%%%%%%%%%%%%%%%%%%%%%%%%%%%%%%%%%%%%%%%%
Semileptonic FCNC
decays $B\to X_s \ell^+ \ell^-$ are known since the
first observation of the $B\to K \ell^+ \ell^-$ decay by Belle~\cite{Ref:Kll-obs}.
Here we report about the first observation of the long awaited
$B\to K^\ast \ell^+ \ell^-$ decay and about a semi-inclusive analysis.

%%%%%%%%%%%%%%%%%%%%%%%%%%%%%%%%%%%%%%%%%%%%%%%%%%%%%%%%%%%%%%%%%%%%%%%%%%%%%%%%
\subsection{First observation of $B\to K^\ast ll$}\label{sec:kll}
%%%%%%%%%%%%%%%%%%%%%%%%%%%%%%%%%%%%%%%%%%%%%%%%%%%%%%%%%%%%%%%%%%%%%%%%%%%%%%%%
This analysis~\cite{Ref:Kll} 
searches for $B\to K^\ast ll$ and $B\to K ll$ using 
the full $140\:\rm fb^{-1}$ data sample available in Summer 2003.
The candidates are formed using an oppositely-charged lepton pair 
(muons or electrons) and a $K^+$, $K^0_S$, or a 
$K^\ast$ candidate formed formed as $K^+\pi^-$, $K_S^0\pi^+$ or
$K^+\pi^0$. The lepton pair is vetoed if its mass 
is below $140\:{\rm MeV}/c^2$, or compatible with the $J/\psi$
or $\psi'$ masses. In the $eeK^\ast$ case, we also consider $ee\gamma$ and 
$ee\gamma\gamma$ combinations to suppress 
the $\psi^{(')}$ background due to Bremsstrahlung. 
%%%%%%%%%%%%%%%%%%%%%%%%%%%%%%%%%%%%%%%%
\begin{figure}
\centerline{\resizebox{0.8\columnwidth}{!}{\includegraphics{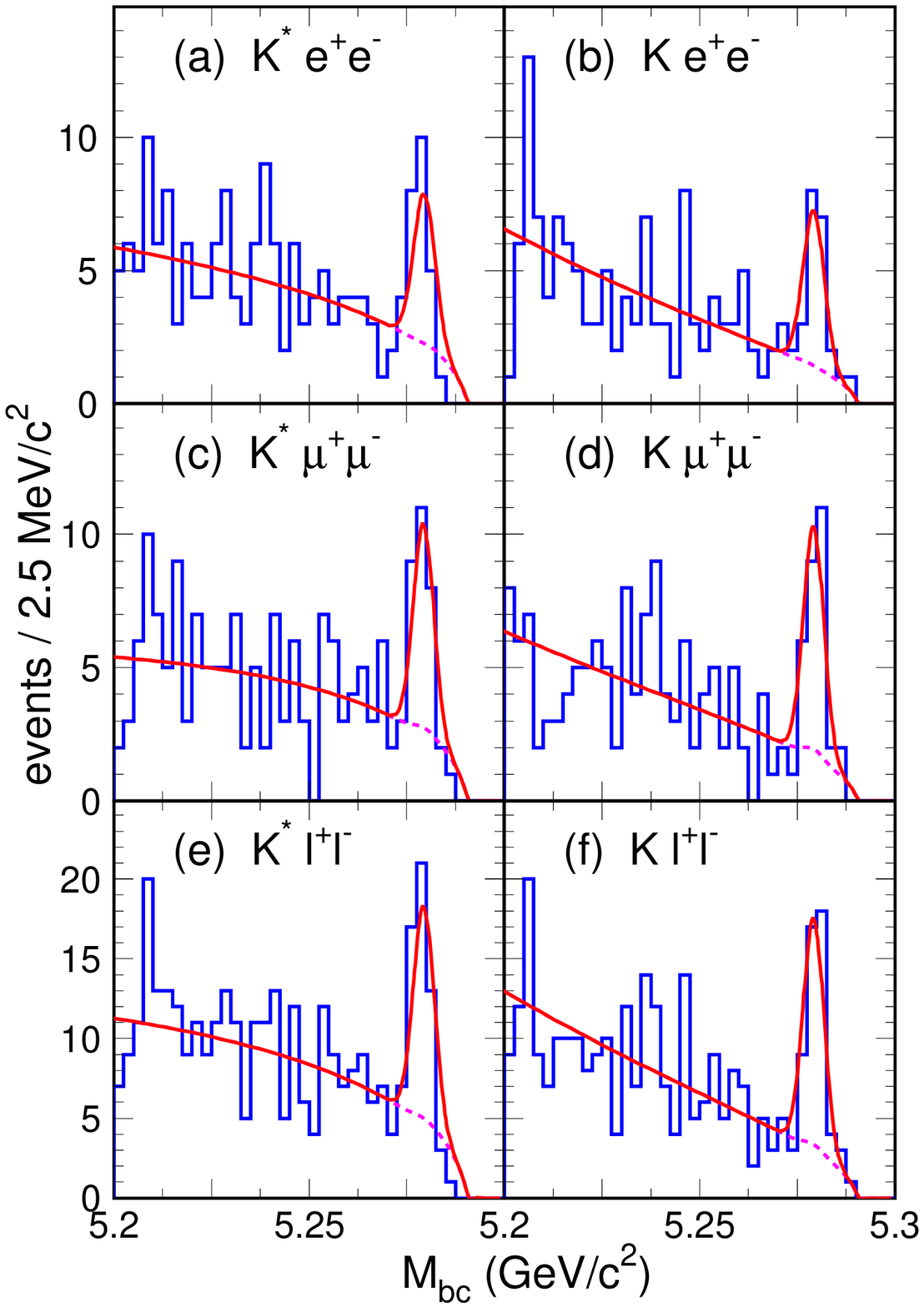}}}
\caption{$m_{\rm bc}$ fits for $K^\ast ll$ and $Kll$ final states.}
\label{fig:Kll}  
\end{figure}
%%%%%%%%%%%%%%%%%%%%%%%%%%%%%%%%%%%%%%%%
The fitted $m_{\rm bc}$
distributions are shown in Fig.~\ref{fig:Kll}. We observe 
$36\pm8$ $B\to K^\ast \ell^+\ell^-$ and $38\pm8$ $B\to K \ell^+\ell^-$ events, 
with statistical significances of $5.7\sigma$ and $7.4\sigma$ respectively
We extract the following preliminary branching fractions:
\begin{eqnarray*}
{\mathcal B}\left(B\to K^{\ast}\ell^+ \ell^-\right) & = & 
  \left(11.5{\:}^{+\:2.6}_{-\:2.4}\pm0.8\pm0.2\right)\cdot 10^{-7} \\
{\mathcal B}\left(B\to K\ell^+ \ell^-\right)        & = & 
  \left(\ \, 4.8{\:}^{+\:1.0}_{-\:0.9}\pm0.3\pm0.1\right)\cdot 10^{-7}
\end{eqnarray*}
where the third error is due to model-dependence.
%%%%%%%%%%%%%%%%%%%%%%%%%%%%%%%%%%%%%%%%
\begin{figure}
\resizebox{\columnwidth}{!}{\includegraphics{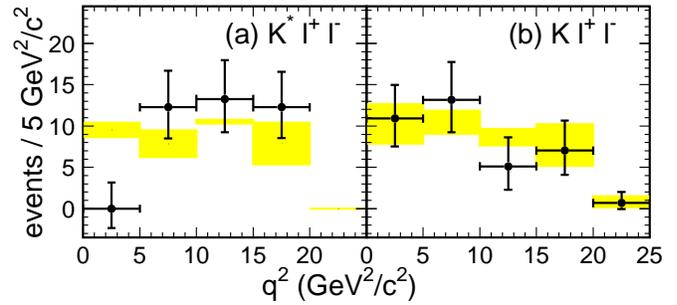}}
\caption{$q^2$ distributions for $Kll$ and $K^\ast ll$.
         Points show data while bands show the expectation range
	 of various models~\cite{Ref:Kll-models}.}
\label{fig:Kll2}  
\end{figure}
%%%%%%%%%%%%%%%%%%%%%%%%%%%%%%%%%%%%%%%%
Fig.~\ref{fig:Kll2} shows the measured squared dilepton mass ($q^2$) distributions
compared to theoretical predictions~\cite{Ref:Kll-models}.

%%%%%%%%%%%%%%%%%%%%%%%%%%%%%%%%%%%%%%%%%%%%%%%%%%%%%%%%%%%%%%%%%%%%%%%%%%%%%%%%
\subsection{Semi-inclusive analysis}\label{sec:xsll}
We performed a semi-inclusive analysis
using $60\:\rm fb^{-1}$~\cite{Ref:Xsll}. In this case the lepton pair is
combined with any of 18 combinations made of one kaon ($K^\pm$ or $K_S^0$)
and up to four pions, one of which may be neutral. The so formed $X_s$
system is required to have a mass below $2.6\:{\rm GeV}/c^2$. 
%%%%%%%%%%%%%%%%%%%%%%%%%%%%%%%%%%%%%%%%
\begin{figure}
\resizebox{\columnwidth}{!}{\includegraphics{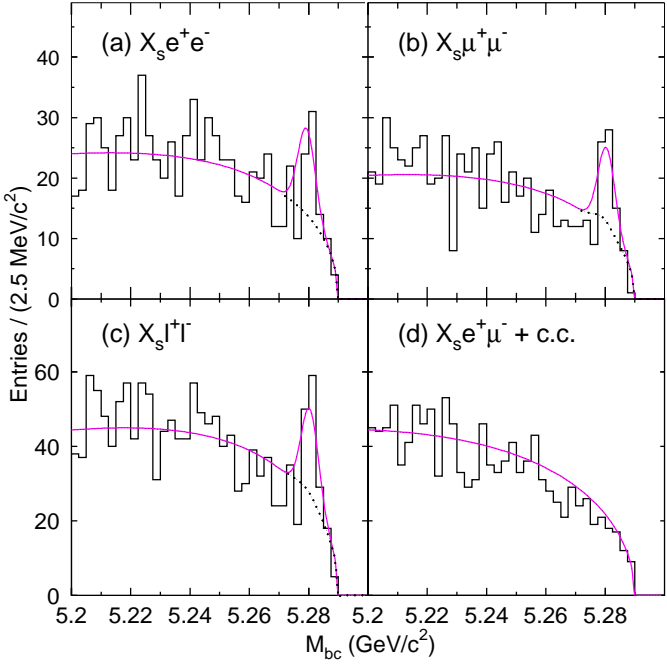}}
\caption{$m_{\rm bc}$ fits for $X_s ll$ final states.}
\label{fig:Xsll1}  
\end{figure}
%%%%%%%%%%%%%%%%%%%%%%%%%%%%%%%%%%%%%%%%
The $m_{\rm bc}$ mass fits are shown in Fig.~\ref{fig:Xsll1} for
$B\to X_s ee$, $B\to X_s \mu\mu$ and the sum $B\to X_s \ell\ell$,
where peaks are seen at the $B$ mass. The forbidden $B\to X_s e\mu$ mode is also
shown as a control sample. We observe $60\pm14{\:}^{+\:9}_{-\:5}$ $B\to X_s \ell\ell$
events with a statistical significance of $5.4\sigma$. The branching fractions
are:
\begin{eqnarray*}
{\mathcal B}\left(B\to X_s \ell^+ \ell^-\right) & = & 
  \left(6.1\pm1.4{\:}^{+\:1.4}_{-\:1.1}\right)\cdot 10^{-6} \quad (5.4\sigma)\\
{\mathcal B}\left(B\to X_s e^+ e^-\right) & = & 
  \left(5.0\pm2.3{\:}^{+\:1.3}_{-\:1.1}\right)\cdot 10^{-6} \quad (3.4\sigma) \\
{\mathcal B}\left(B\to X_s \mu^+ \mu^-\right) & = & 
  \left(7.9\pm2.1{\:}^{+\:2.1}_{-\:1.5}\right)\cdot 10^{-6} \quad (4.7\sigma).
\end{eqnarray*}

%%%%%%%%%%%%%%%%%%%%%%%%%%%%%%%%%%%%%%%%
%\begin{figure}
%\resizebox{\columnwidth}{!}{\includegraphics{fig1_xs-ll-color.eps}}
%\caption{$m_{ll}$ (left) and $m_{X_s}$ (right) MC (top) and 
%         measured (bottom). The efficiency corrected distribution is shown
%	 for comparison.}
%\label{fig:Xsll2}  
%\end{figure}
%%%%%%%%%%%%%%%%%%%%%%%%%%%%%%%%%%%%%%%%
%Fig.~\ref{fig:Xsll2}~\cite{Ref:Xsll}

%%%%%%%%%%%%%%%%%%%%%%%%%%%%%%%%%%%%%%%%%%%%%%%%%%%%%%%%%%%%%%%%%%%%%%%%%%%%%%%%
%
\section{Leptonic FCNC $B$ decays}\label{sec:ll}
%
%%%%%%%%%%%%%%%%%%%%%%%%%%%%%%%%%%%%%%%%%%%%%%%%%%%%%%%%%%%%%%%%%%%%%%%%%%%%%%%%
Finally, we report about the search 
for the FCNC decays $B\to ee$ $B\to \mu\mu$ and 
$B\to e\mu$, using a data sample of $78\:\rm fb^{-1}$~\cite{Ref:ll}.
The Standard Model (SM) branching fractions predictions for the first two decays
are about $10^{-10}$ and $10^{-15}$ respectively, but they could be enhanced by two order 
of magnitude in models including two Higgs doublets or $Z$-mediated FCNC.
Apart from the negligibly small contribution form neutrino oscillations,
the $B\to e\mu$ is forbidden in the SM, but could occur in some SUSY models
or the Pati-Salam leptoquark model~\cite{Ref:Pati-Salam}.

The selection is based on stringent requirements for the particle-identification
of the two leptons and strong requirements for the $q\bar{q}$ ($q=u,d,s,c$)
and $\tau\tau$ background rejections. In particular, to favour $B\bar{B}$ events, 
we require the presence of five charged tracks in the event.

\begin{figure}
\resizebox{0.32\columnwidth}{!}{\includegraphics{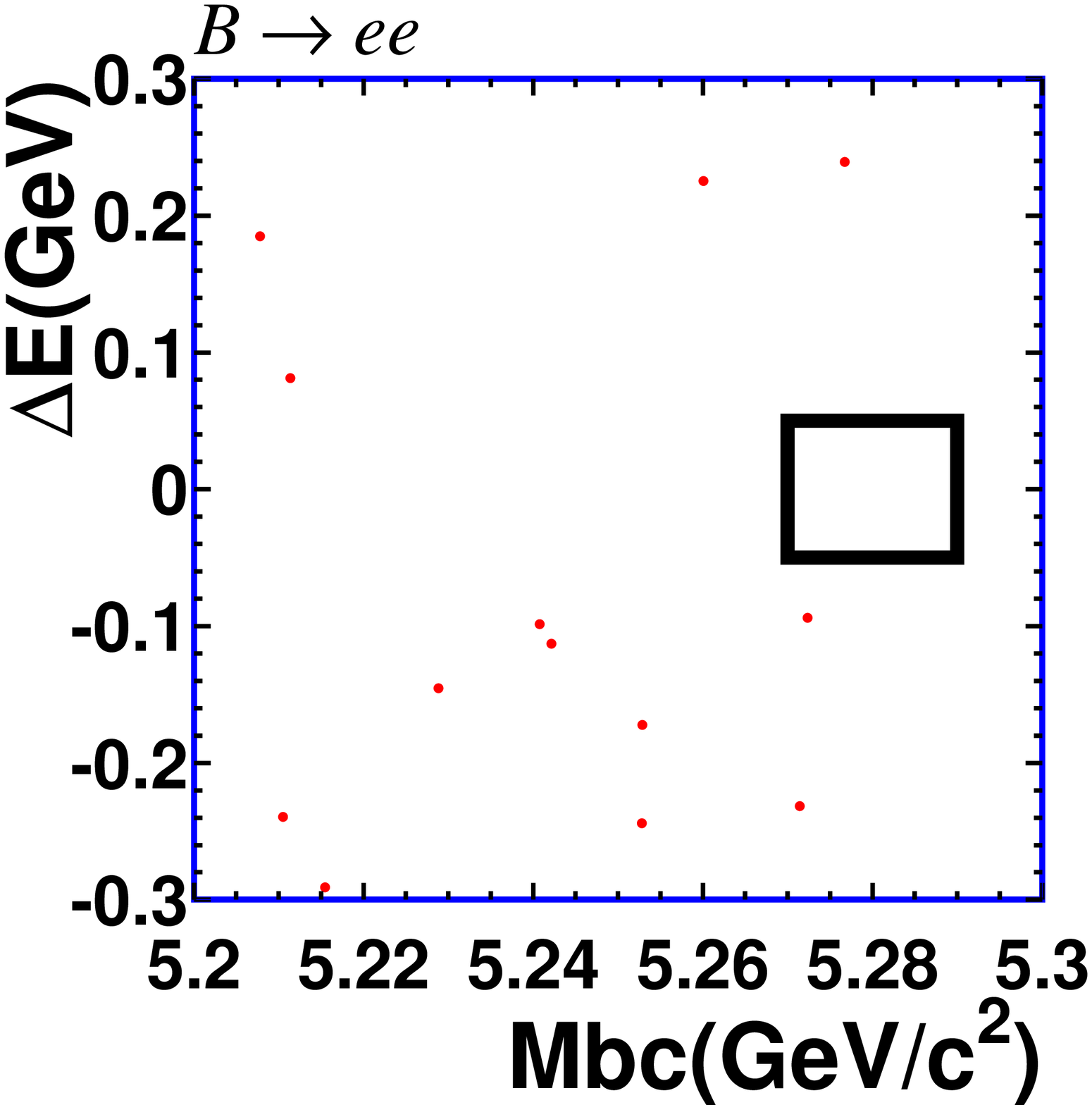}}
\resizebox{0.32\columnwidth}{!}{\includegraphics{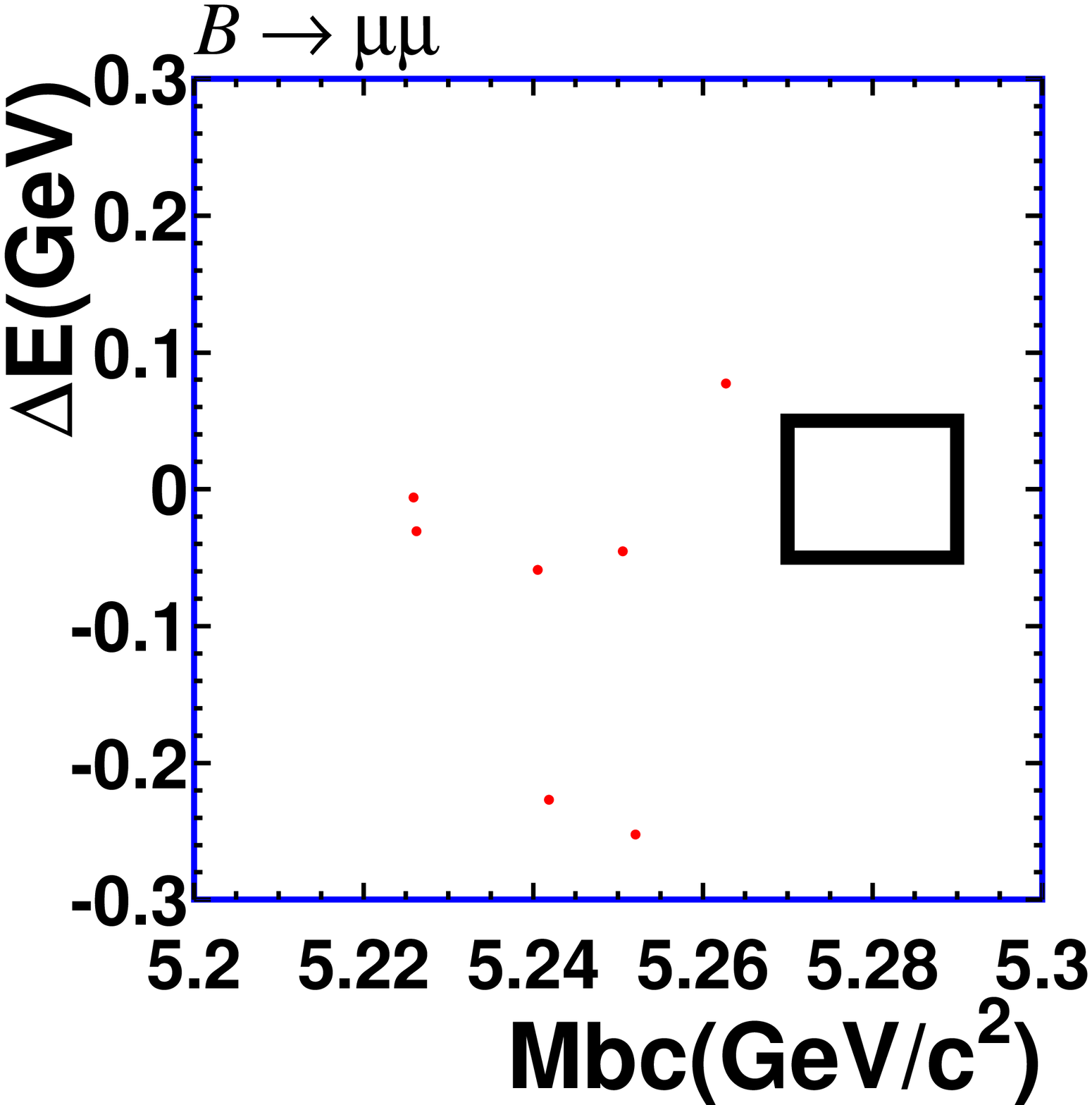}}
\resizebox{0.32\columnwidth}{!}{\includegraphics{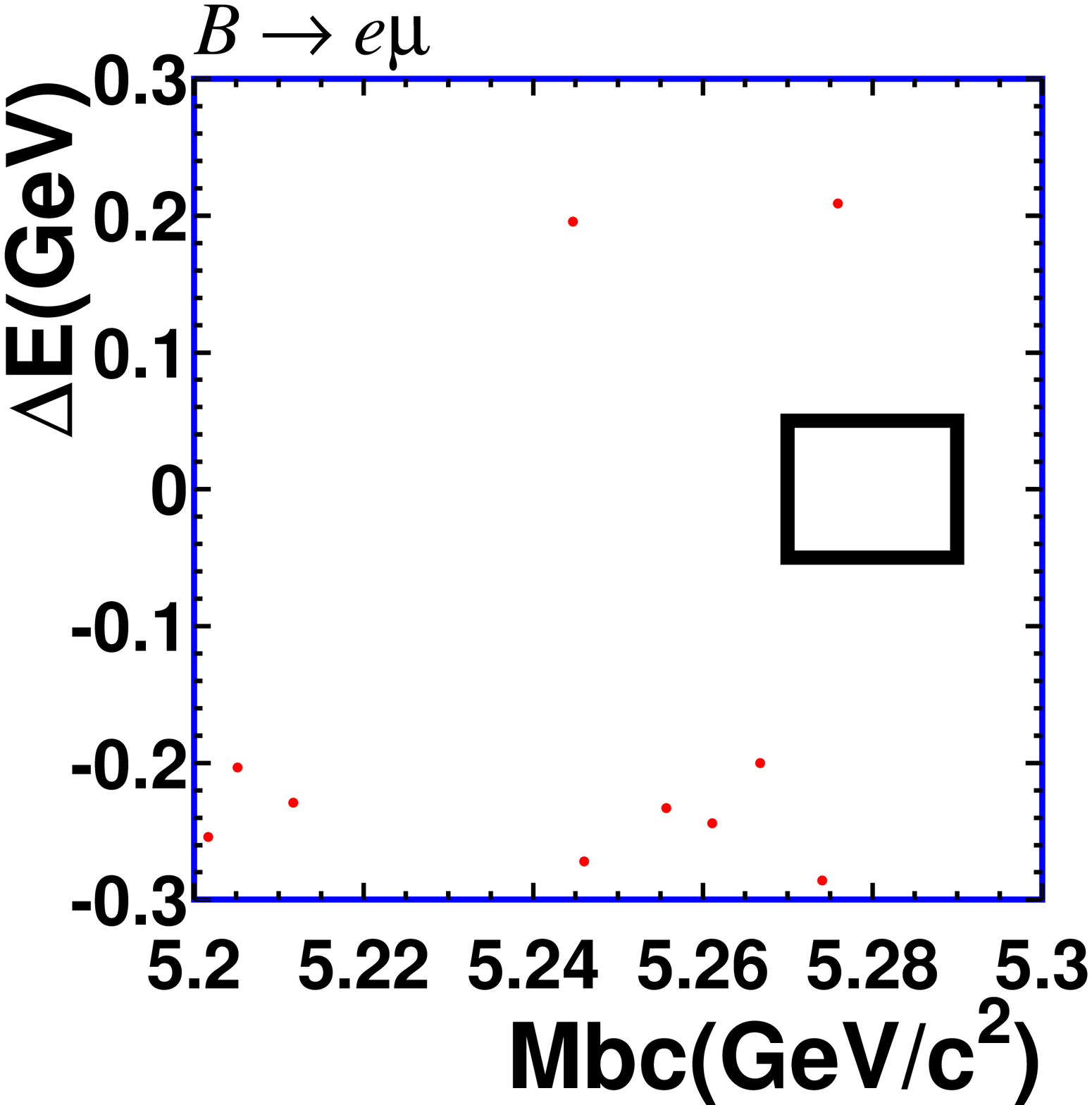}}
\caption{$\Delta E$ versus $m_{bc}$ for $ee$, $\mu\mu$ and $e\mu$ final states.
The rectangles indicate the signal box.}
\label{fig:ll}  
\end{figure}

We find no events in the signal box defined in the $\Delta E$--$m_{\rm bc}$ 
plane, as shown in Fig.~\ref{fig:ll}, 
%%%%%%%%%%%%%%%%%%%%%%%%%%%%%%%%%%%%%%%%
%%%%%%%%%%%%%%%%%%%%%%%%%%%%%%%%%%%%%%%%
while we expect about $0.2$ to $0.3$
events from background, depending on the mode. 
We set upper limits on the branching fractions as:
\begin{eqnarray*}
{\mathcal B}\left(e^+e^-\right) & < & 1.9 \cdot 10^{-7} \quad (90\%\:{\rm CL})\\
{\mathcal B}\left(\mu^+\mu^-\right) & < & 1.6 \cdot 10^{-7} \quad (90\%\:{\rm CL}) \\
{\mathcal B}\left(e^\pm\mu^\pm\right) & < & 1.7 \cdot 10^{-7} \quad (90\%\:{\rm CL}). 
\end{eqnarray*}
The latter allows to set a 90\% CL lower limit on the mass of the 
Pati-Salam leptoquark~\cite{Ref:Pati-Salam,Ref:Pati-Salam2} at
$46\:{\rm TeV}/c^2$. The details of the extraction are given in
Ref.~\cite{Ref:ll}.

%%%%%%%%%%%%%%%%%%%%%%%%%%%%%%%%%%%%%%%%
\section{Conclusion}
While radiative $B$ decays become tools to understand the
QCD structure of the $B$ meson, semileptonic FCNC
decays become hot candidates to test
extensions of the Standard Model. After a long wait, we finally 
observed the decay $B\to K^\ast\ell\ell$, opening the road
to measurements of the lepton forward-backward asymmetry.

\end{document}